\documentstyle[twocolumn,aps]{revtex}
\begin{document}
\draft
\title{Paramagnetic Meissner effect in mesoscopic samples.} 
\author
{V.A. Schweigert $^{\dag}$ and F.M. Peeters $^{\ddag}$}
\address{
$^{\dag}$ Institute of Theoretical and Applied Mechanics,
Novosibirsk, Russia,\\ 
$^{\ddag}$ Departement Natuurkunde, Universiteit Antwerpen (UIA), B-2610 Antwerpen,
Belgium \\
\tt Electronic mail: peeters@uia.ua.ac.be}
\date{\today}
\maketitle
\begin{abstract}
Using the non-linear Ginzburg--Landau (GL) theory, we 
study the magnetic response of different shaped samples 
in the field--cooled regime (FC).
For high external magnetic fluxes,
the conventional diamagnetic response under cooling down 
can be followed by the paramagnetic Meissner effect (PME). 
A second-order transition from
a giant vortex state to a multi--vortex state, with the same vorticity,
occurs at the second critical field which leads to the suppression of PME.
\end{abstract}

The Meissner effect is considered the most important characteristic 
property of superconductivity. When a superconductor is
cooled down in the presence of an external magnetic field the  
field is expelled and it behaves
as a diamagnet. However, some samples show a paramagnetic response
under cooling. The finding of PME (or Wohlleben effect) in high-$T_c$
superconductors \cite{braunisch} initiated the appearance of several models
interpreting PME as evidence of  non-conventional superconductivity
in these materials (e.g. see Ref.\cite{sigrist}).
However, numerous observations of PME in conventional macroscopic
\cite{crus,thompson,kostic,pust,terentiev} and mesoscopic \cite{geim}
superconductors indicate the existence of another mechanism, which may be 
explained with GL theory.

Based on results from axial-symmetric solutions of the GL equations
 by Fink and Presson \cite{fink}, 
Cruz {\it et al.} \cite{crus} proposed that PME in their experiments
on $Pb_{99}Tl_{01}$ cylinders is caused by a temperature variation
of the superconducting density in a giant vortex state with
 fixed angular momentum \cite{crus}. In such a state, the superconducting
current, which shields the magnetic field in the vicinity of the
sample boundary (essentially the Meissner effect), changes its direction
expelling magnetic field {\it into} the sample that can lead to
the PME. Thereafter this idea, often called {\it
flux compression}, was exploited in \cite{geim,moshchalkov,koshelev},
but a quantitative analysis of PME within GL theory is still missing
and several principal questions
remain to be answered: 1) is the vorticity of 
the giant vortex state fixed
during cooling down as was assumed in Refs.~\cite{crus,moshchalkov}?,
 2) if yes, can this lead to
the appearance of PME?, and, finally, 3) can the proposed mechanism
explain  the PME in recent experiments with conventional macroscopic
\cite{crus,thompson,kostic,pust,terentiev} and 
mesoscopic \cite{geim} samples? In this Letter, we follow the
GL approach and address these
questions by studying the magnetic response of different--shaped samples
in the FC regime.

We consider a defect--free 
superconducting disk (and cylinder) immersed in an insulating media
with a perpendicular (along cylinder axis) uniform magnetic field $H_0$.
The behaviour of the superconductor is characterized
by its radius $R$ (and thickness $d$ for the disk case), magnetic
field $H_0$, the coherence $\xi(T)=\xi(0)(1-T/T_c)^{-1/2}$ and penetration
$\lambda(T)=\lambda(0)(1-T/T_c)^{-1/2}$ lengths, where
$T_c$ is the critical temperature.
 To reduce the number of independent variables we measure
the distance in  units of the sample radius, the vector
potential $\vec A$ in $c\hbar/2eR$, and the order parameter $\Psi$
in $\sqrt{-\alpha/\beta}$ with $\alpha$, $\beta$ being the GL
coefficients \cite{gennes}. Then the GL equations become 
\begin{equation}
\label{eq1}
\left(-i\vec \nabla_{2D} -\vec A\right)^2\Psi=\frac{R^2}{\xi^2}
\Psi (1-|\Psi|^2),
\end{equation}
\begin{equation}
\label{eq2}
-\triangle_{3D} \vec A=\frac{R^2}{\lambda^2} f(z)\vec j_{2D}.
\end{equation}
Here,  the indices $2D$, $3D$ refer to two-dimensional 
and three-dimensional operators;
\begin{equation}
\label{eq4}
\vec j_{2D}=\frac{1}{2i}
\left(\Psi^*\vec \nabla_{2D} \Psi-\Psi \vec\nabla_{2D}\Psi^*\right)
-|\Psi |^2\vec A, 
\end{equation}
is the density of superconducting current in
the plane ($x,y$), and the external magnetic field is directed along
the $z$-axis. The boundary conditions to Eqs.~(\ref{eq1}-\ref{eq2})
correspond to zero superconducting current at the sample boundary
and uniform magnetic field far from the sample.
The order parameter and superconducting current
are assumed to be independent of 
the $z-$coordinate which is valid for cylinders as well as for thin
$d\ll \xi,\lambda$ disks \cite{deo,schweigert,schweigert2}. Then $f(z)=1$ and
$f(z)=d\delta(z)$ for the cylindrical and disk geometry, respectively.

Superconducting disks and cylinders placed in a magnetic field and
cooled down transit from a normal 
to a superconducting state at the critical temperature $T_{\star}$,
which depends both on $H_0$ and $R$. However, the unitless parameters
$H_0/H_{c2}(T_{\star})$, $R/\xi(T_{\star})$, and the angular momentum 
of the giant vortex superconducting state in the nucleation point $T_{\star}$
depend  only on the magnetic flux
$\Phi=\pi R^2H_0$ piercing through the sample \cite{schweigert}. Here, 
$H_{c2}=\Phi_0/2\pi\xi^2$ and $\Phi_0=hc/2e$ are the
second critical field and the flux quantum, respectively. With further
cooling down, the magnetic response is characterized by only
two independent variables $H_0/H_{c2}$ and $\Phi$.

Our numerical approach for solving Eqs.~(\ref{eq1}-\ref{eq2}) is
described in \cite{deo,schweigert,schweigert2}. It turns out that
an accurate simulation of a multi-vortex state is a hard task in the
case of
large vorticity $L$. The latter correponds to the total angular momentum and
the number of vortices in the giant-vortex and the multi--vortex state,
respectively.
To improve the accuracy we apply a non-uniform
rectangular space grid condensing in the vicinity of the sample boundary.
However, due to
the tremendous computational expenses (e.g.
the total number of grid points in the plane ($x,y$) was
about 160000 for $L=30$) we could only treat 
the vortex state with large $L$ 
in  the cylindrical case, when the vector
potential is uniform in the $z$-direction.
In the disk case, we restrict our considerations 
to the axial symmetric solutions
$\Psi=\psi(\rho)exp(iL\phi)$ ($\rho$, $\phi$
are the cylindrical coordinates), which are shown to be stable in the
region $H_0>H_{c2}$.

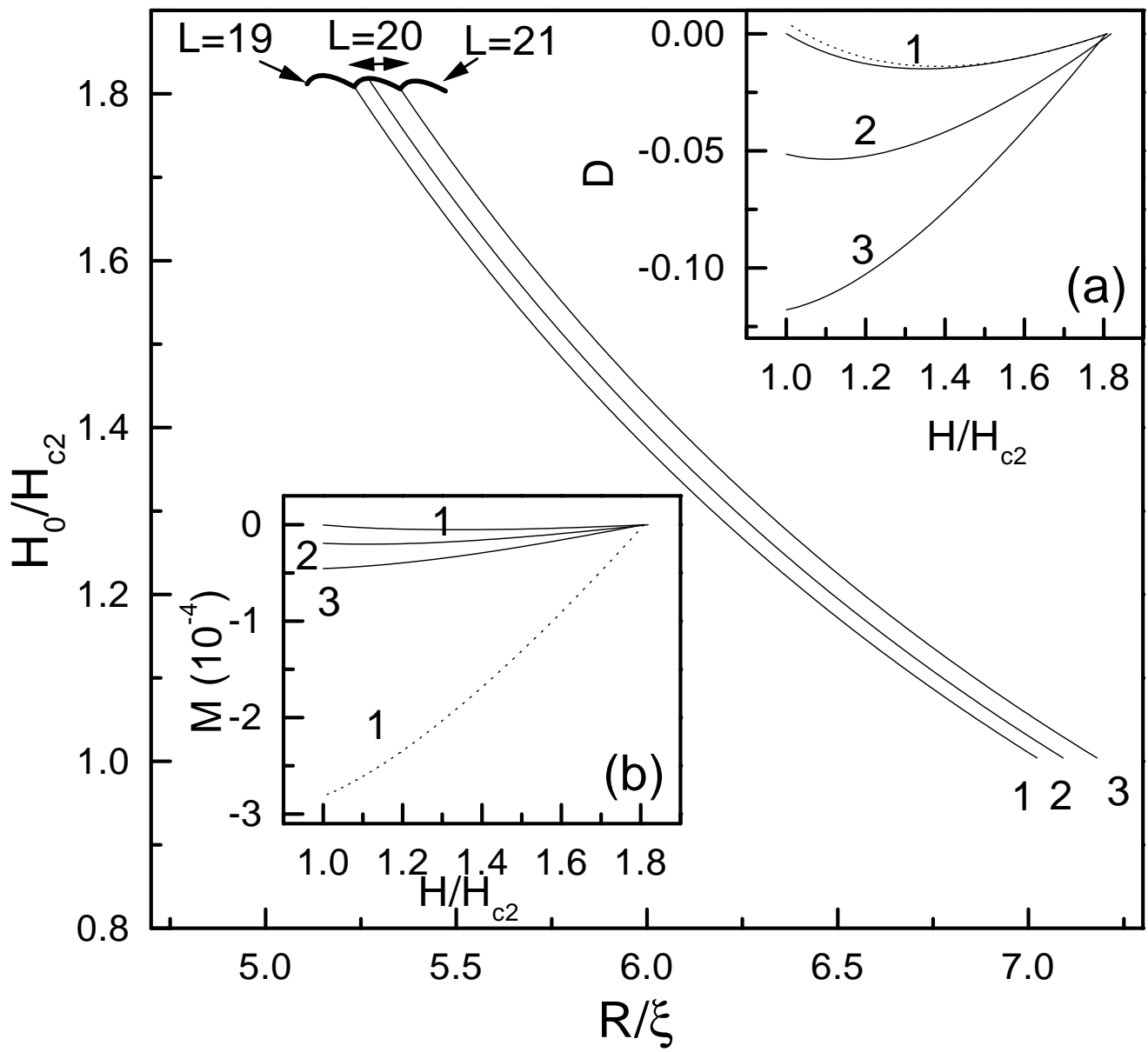
\begin{figure}
\caption{
The nucleation field as a function of the radius. Thin curves
show the path along which the system evolves during cooling down.
Insets (a) and (b) show the
corresponding dipole magnetic moment and magnetization
($M=H_{c2}^{-1}\int d\vec r(H_z-H_0)/d\vec r$, where integration is
performed over the sample volume), respectively.
Here, solid and dotted curves correspond to cylinder ($\kappa=10$) and 
disk ($\kappa=1$, $d=0.1R$), respectively.
}
\label{fig1}
\end{figure}

When the sample is cooled down below the critical temperature $T_{\star}$, 
a giant vortex state appears with 
angular momentum $L$, which is determined by the magnetic flux
$\Phi$ \cite{moshchalkov,schweigert,benoist}. Starting from this
state we mimick the FC regime by decreasing (increasing) slowly the value
of $H_0/H_{c2}\sim 1/(1-T/T_c)$ ($R^2/\xi^2\sim (1-T/T_c)$) such that the system
evolves along a path with fixed external magnetic flux (e.g. see Fig.~\ref{fig1}).
Using the superconducting state found at the previous step as input, we find
the next steady--state solution to Eqs.~(\ref{eq1}-\ref{eq2}).
Doing so we consider only stable solutions and neglect thermal
fluctuations, which could lead to possible transitions between metastable
states. This assumption is valid for normal superconductors where
the barriers separating the metastable states exceed by far the sample temperature,
except near points in which the state becomes unstable \cite{schweigert3},
e.g. near the saddle points. 

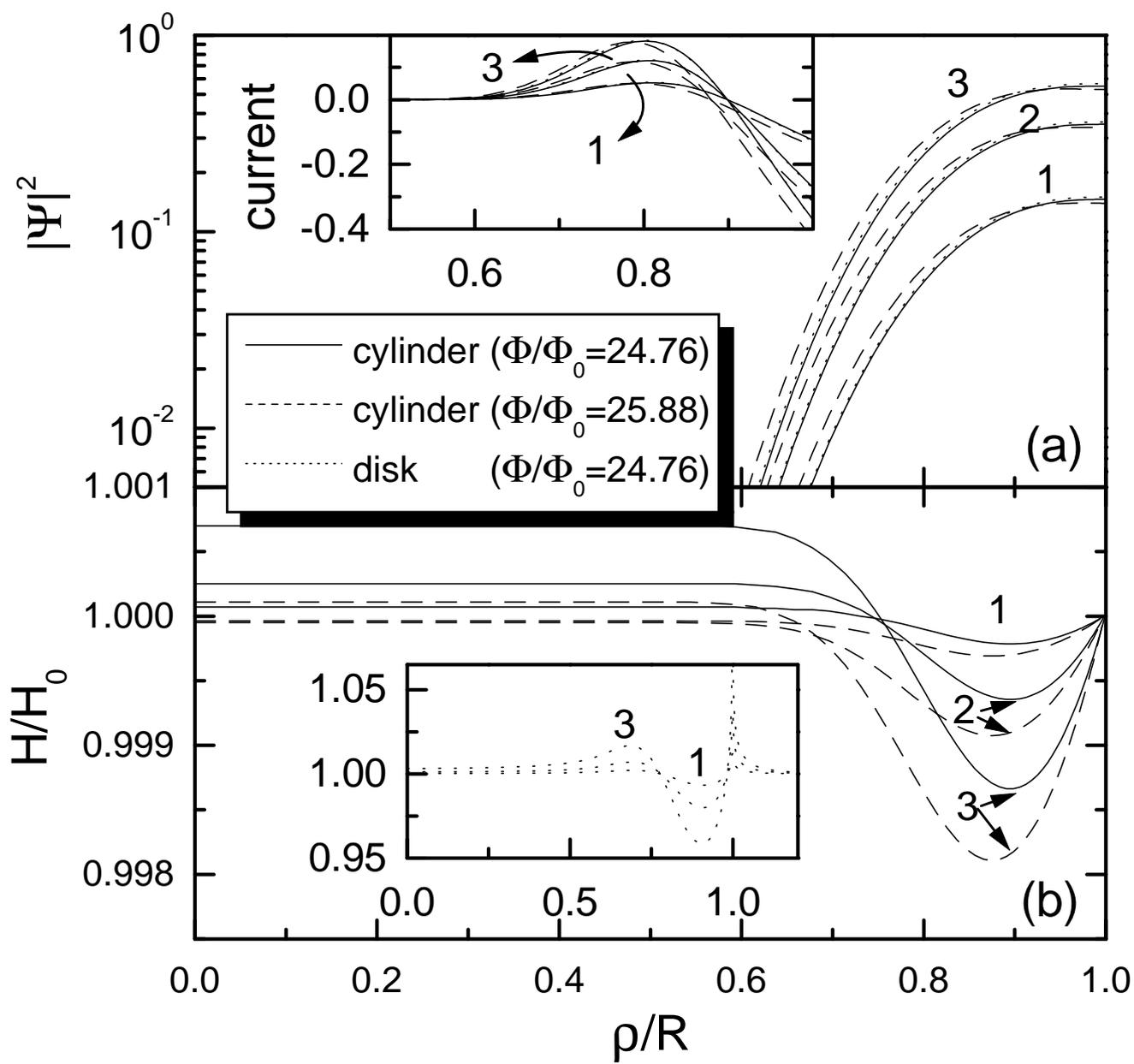
\begin{figure}
\caption{
The distribution of the superconducting electrons density (a), the
superconducting current (inset) and the magnetic field (b) 
in the cylinder (solid and dashed curves) and the disk (inset)
for different reduced magnetic fields $H/H_{c2}:$ 1-1.6, 2-1.3, 3-1.0.   
}
\label{fig2}
\end{figure}

When calculating the dipole magnetic moment,
we can neglect non-linear effects in the vicinity of the nucleation point. 
The quantum angular momentum $L$ increases almost proportional
to the magnetic flux $\Phi$ but remains always smaller
than $\Phi/\Phi_0$. The supervelocity
$v_{\phi}=\rho^{-1}(L-\Phi\rho^2/\Phi_0R^2)$, which is
oriented along the azimuthal direction,
changes its sign at $\rho_{\star}=R\sqrt{L\Phi_0/\Phi}$. Therefore, 
both diamagnetic ($\rho>\rho_{\star}$) and paramagnetic
($\rho<\rho_{\star}$) currents exist in any giant vortex state.
However, the magnetic moment, which
can be estimated in the lowest Landau level (LLL) approximation as 
$D\sim \int d\rho \rho^2v_{\phi}|\psi_L|^2$ with $\psi_L$
being the lowest eigenfunction of the linearized first GL equation,
turns  out to be always diamagnetic both for disks and cylinders. As long as
the superconducting density $|\Psi|^2$ remains small,
the magnetic moment almost linearly increases in absolute value
with decreasing $H_0/H_{c2}$ (inset (a) in Fig.~\ref{fig1}).
With further cooling down, the LLL approximation
breaks down. Due to non-linear effects 
(mainly from the second term in the RHS of Eq.~(\ref{eq1}))
the order parameter increases more rapidly in the inner region 
which leads to  an increase of the paramagnetic component 
(Figs.~\ref{fig1},\ref{fig2}).

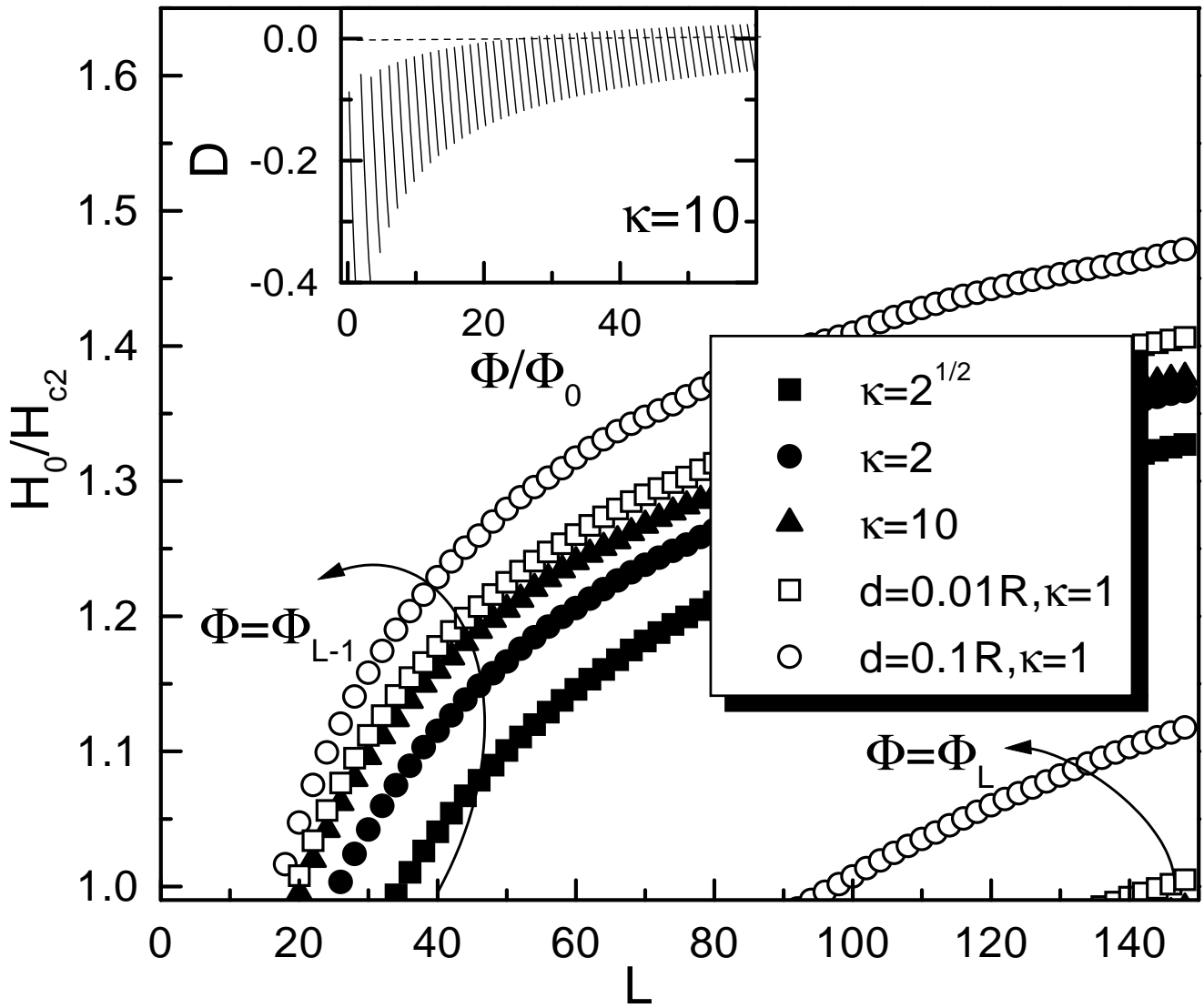
\begin{figure}
\caption{The magnetic field corresponding to the transition
from the diamagnetic to paramagnetic response as a function of the total
angular momentum for $\Phi=\Phi_{L-1}$ and $\Phi=\Phi_{L}$.
Solid and open symbols correspond to cylinders and disks, respectively. Inset
shows the magnetic moment of the cylindrical sample at $H_0=H_{c2}$ and
$\kappa=10$.}
\label{fig3}\end{figure}

The resulting magnetic moment crucially depends on 
the ratio between  $L$ and $\Phi/\Phi_0$.
With increasing magnetic field, the switching $L\rightarrow L+1$
of the angular momentum of the nucleated state occurs
at a certain $\Phi_L$ \cite{schweigert}.
For given $L$, the magnetic moment reaches its maximum and minimum value at 
$\Phi_{L-1}$ and $\Phi_{L}$, respectively. 
Due to angular momentum quantization the magnetic moment
exhibits a strong oscillating behaviour as function of the magnetic field 
(see inset of Fig.~\ref{fig3}) which agrees with Geim's
observations \cite{geim}. 
With decreasing temperature, the dipole magnetic moment becomes zero 
for certain ratio $H/H_{c2} \sim 1/(1-T/T_c)$ and the original
diamagnetic response becomes paramagnetic. 
This magnetic field is shown in Fig.~\ref{fig3} for 
$\Phi=\Phi_{L-1}$ and $\Phi=\Phi_{L}$. Note that: 1) for the cylinder geometry
a larger GL parameter $\kappa=\lambda/\xi$ favours PME, 
2) while an increase of
the effective penetration length $\lambda^2/d$ suppresses PME in disks,
and 3) an increase of $\Phi$ favours PME both in cylinders and disks. 
The reason is that 
the point $\rho_{\star}$, where the supervelocity changes its direction,
shifts towards the sample boundary with increasing angular momentum which 
is related to the total magnetic flux $\Phi(L\gg 1)\approx \Phi_0(L+L^{1/2})$
\cite{benoist}.  

In cylinders, the magnetization is directly proportional to the magnetic moment.
In disks: 1) the diamagnetic currents flowing near the sample boundary
give larger contributions to the magnetization than inner paramagnetic currents
\cite{koshelev} and consequently, the paramagnetic contribution to the
magnetization will be strongly suppressed (inset (b) in Fig.~\ref{fig1}),
2) but the smaller trapped magnetic flux, in disks as compared to cylinders,
decreases the diamagnetic response.
This is the reason why thicker disks show onset of PME at larger
 $H_0/H_{c2}$ (Fig.~\ref{fig3}, open symbols).

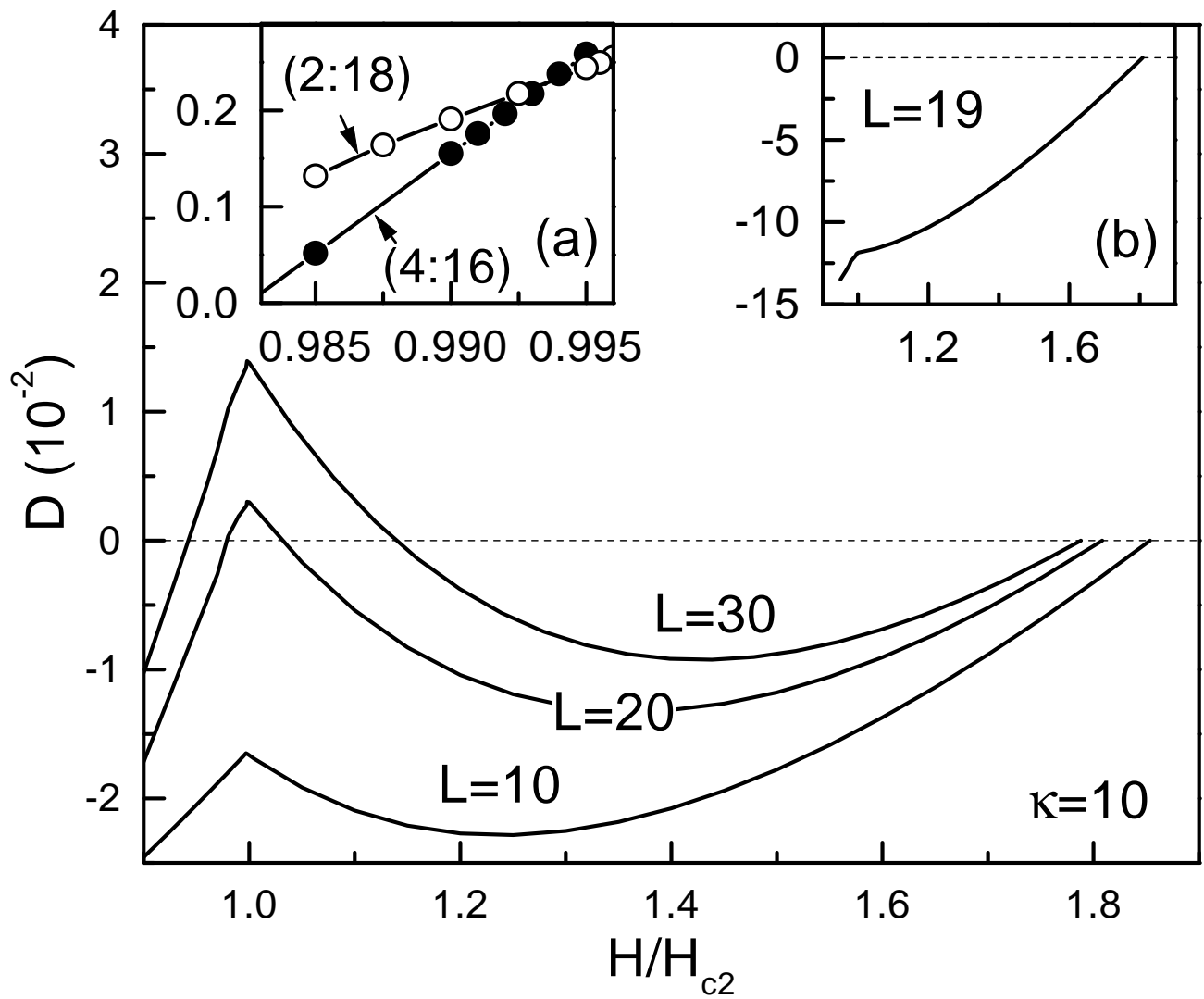
\begin{figure}
\caption{
The dipole magnetic moment of  cylindrical samples for different
external magnetic fluxes $\Phi=\Phi_{L-1}$ and $\kappa=10$. 
The hysteretic behaviour is demonstrated 
in inset (a) for $L=20$ and $\Phi=\Phi_{L-1}$.
Inset (b) shows the magnetic response for $L=19$ and $\Phi_{L}$.
}
\label{fig4}
\end{figure}

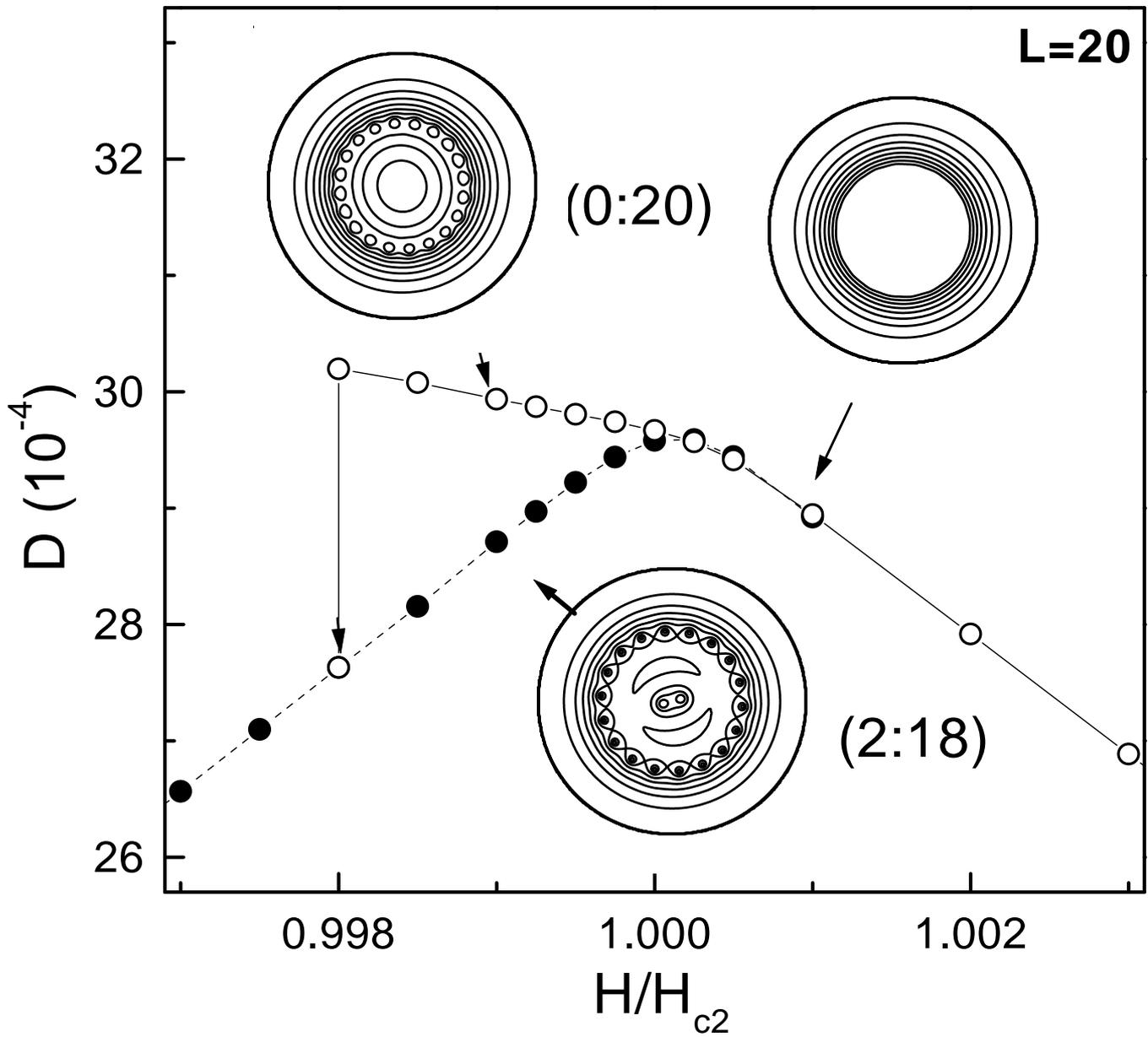
\begin{figure}
\caption{
The dipole magnetic moment and contour plots of the superconducting 
density (log-scale is used) in the vicinity of the second critical field.
}
\label{fig5}
\end{figure}

When the second critical field $H_{c2}(T)$ becomes smaller than the applied
magnetic field, the giant--vortex state transits to the multi--vortex
state with the same vorticity (Figs.~\ref{fig4},\ref{fig5}).
This {\it second-order transition} is not
followed by any jumps in the magnetization or the magnetic moment.
Just after the transition, all vortices are arranged in a ring (0:L).
Note, that
the magnetic moment of the state (0:L) 
continues to increase with decreasing $H_0/H_{c2}(T)$
but with a smaller slope (Fig.~\ref{fig5}).
With further decreasing $H_0/H_{c2}(T)$ a pair of vortices moves to 
the inner region and the state (2:L-2) appears for $L=20$
(see Fig.~\ref{fig5}).
This {\it first-order transition} is followed by a weak
jump in the magnetic moment. The
derivative $dD/dT$ changes sign and further cooling down
results in the disappearance of PME. 
The magnetic  moment of the diamagnetic state with
smaller angular momentum
$L=19$ and $\Phi=\Phi_{L}$ is also affected
by the transition to the multi--vortex state, which increases the
diamagentic response (inset (b) in Fig.~\ref{fig4}). 
As the temperature decreases, vortices continue to move from the
outer to the inner shell. 
Note, that the corresponding weak jumps in the magnetic moment are
practically  not visible on the  scale used in Fig.~\ref{fig4}.
Although the state with $L=19$ is energetically more favourable
than the one with $L=20$, no vortex exits the system which is 
in agreement with observations \cite{geim}. Since the vorticity
remains unchanged under cooling down, the magnetic response will be
 almost reversible. The hysteresis caused by the 
first--order transitions between different multi--vortex states with
the same vorticity is weak (see inset (a) of Fig.~\ref{fig4}).
Starting from the point $H/H_{c2}=0.9$ and
warming up the system to $H/H_{c2}=1.1$ we find the magnetic moment,
which coincides in the scale of Fig.~\ref{fig4}
with that obtained by cooling down.

Although the magnetization curves, shown in Fig.~\ref{fig4}, 
agree qualitatively with those
from experiments \cite{thompson,kostic,pust}, a number of issues
remain unclear: 1) in experimental observations of macroscopic samples   
a weak hysteresis is found in cooling down and subsequent warming up, and 
2) the observed maximum in the magnetic momentum are
weaker than those from our calculations which 
may be due to the presence of vortex pinning centra.
Note that within GL theory we found a weak PME which is caused by
the competition of large diamagnetic and paramagnetic responses of the
outer and inner part of the sample, respectively. 
Any mechanism which slightly influences any of the two responses 
may strongly influence the total magnetic behavior. As an example, in
macroscopic disks PME disappears after mechanical
abrading the top and bottom surfaces \cite{thompson,kostic,pust}.
A number of samples made
from the same material as those demonstrating PME exhibited
only diamagnetic behaviour \cite{thompson,kostic,pust}. This indicates the
important role played by the sample structural inhomogeneity.
For mesoscopic disks, experiments \cite{geim} show PME for rather small 
angular momenta which does not agree with our simulations for flat circular
mesoscopic disks.

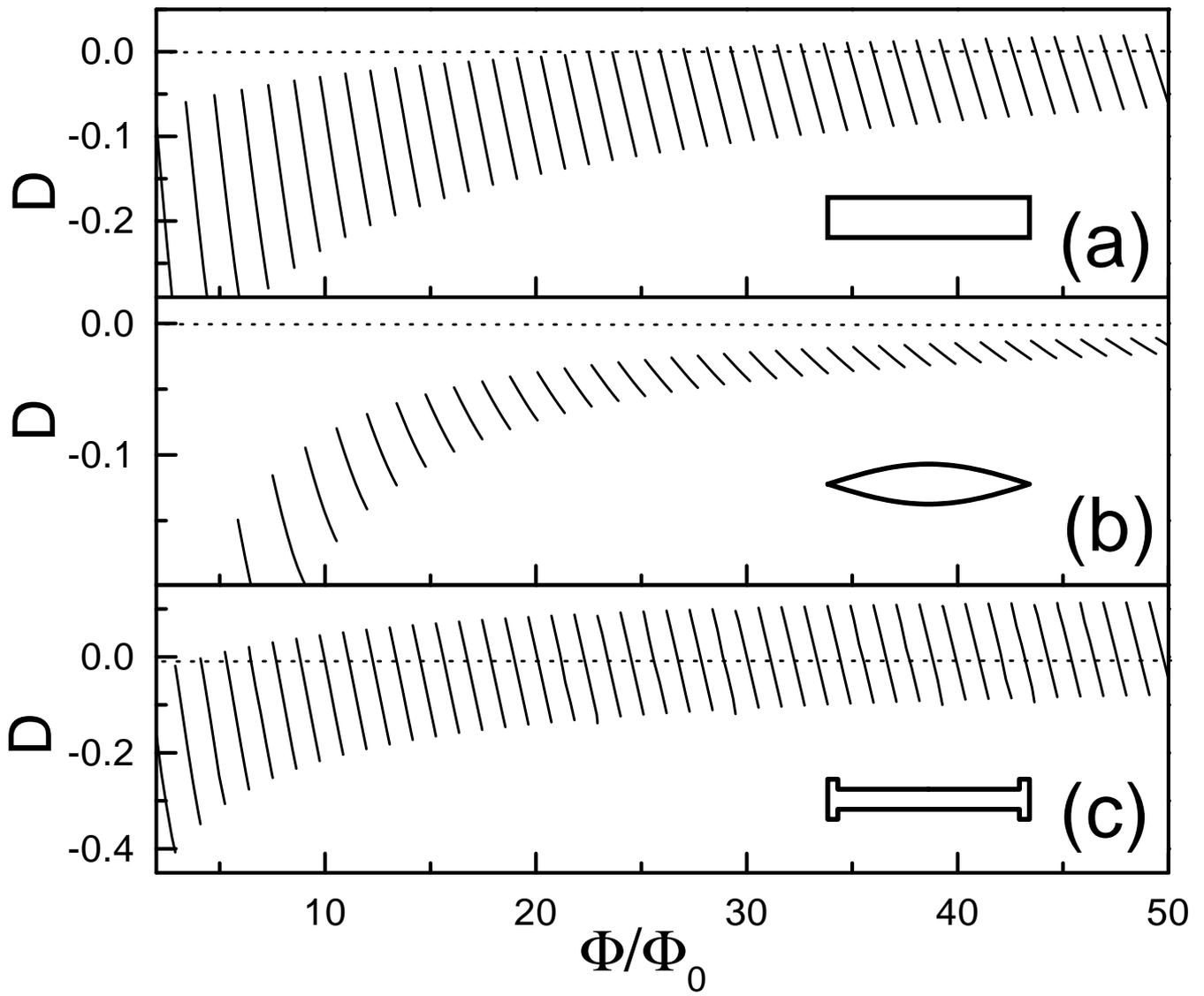
\begin{figure}
\caption{
The dipole magnetic moment of different shaped circular samples
at $H_0/H_{c2}=1$. 
}
\label{fig6}
\end{figure}

To address some of these sample structural issues we consider the influence of
the superconductor shape on the magnetic moment by varying radially the thickness
of the disk. We limit ourselves to the case
of a strong type-$II$ superconductor ($\kappa\gg 1$) and solve, therefore,
only Eq.~(\ref{eq1}). The dipole magnetic moment of different-shaped
samples is shown in Fig.~(\ref{fig6}) for $H_{0}=H_{c2}(T)$, where the
scale of the thickness variation is apparent from the insets of
Fig.~\ref{fig6}. The
magnetic moment becomes more negative (positive) in magnifying
glass (crown) like samples as compared to flat disks.
An increase of the local thickness near the sample boundary (see Fig.~\ref{fig6}(c))
increases the PME. This strongly suggests a possible non flat geometry
of the disks of Ref. \cite{geim}.

In summary, the giant--vortex state remains stable under cooling down
and transits to the multi--vortex state with the same vorticity at
$H_0\approx H_{c2}$. 
The paramagnetic response is caused by a more rapid growth of the
superconducting electron
density in the inner region of the sample, due to non-linear effects,
where paramagnetic currents flow. The appearance of the multi--vortex state
supresses PME and the maximum of the paramagnetic response
corresponds to $H_0\approx H_{c2}$. We showed that within the GL theory 
a paramagnetic response is possible for large magnetic fluxes, which is
in agreement with experimental findings on disks with large radia,
but does not agree with the experimental results on mesoscopic disks
of Geim {\it et al} \cite{geim}.  

After finishing this work, we came aware of a preprint by 
Palacios \cite{palacios00} who used the LLL approximation
to study the FC regime and failed to find any paramagnetic response.
As shown above, one has to go beyond the LLL approximation in order to 
find PME which is caused by non-linear effects. 

We thank A.K.~Geim for useful discussions.
This work is supported by the Flemish Science Foundation (FWO-Vl) and the
``Interuniversity Poles of Attraction Program - Belgian State, Prime Minister's
Office - Federal Office for Scientific, Technical and Cultural Affairs''.
One of us (VAS) was supported by a DWTC fellowship and 
(FMP) is a research director with the FWO-Vl.

\end{document}